\documentclass[a4paper,9pt]{manuscript} 

\usepackage[T1]{fontenc}
\usepackage{mathrsfs}  

\newcommand{\avg}[1]{\E{#1}}
\newcommand{\Var}[1]{\text{Var}[#1]}
\newcommand{\diff}{\mathop{}\!\mathrm{d}}
\newcommand{\deriv}[1]{\frac{\diff{}}{\diff{#1}}}

\newcommand{\E}[1]{\mathbb{E}\left[#1\right]}
\newcommand\varpm{\mathbin{\vcenter{\hbox{%
				\oalign{\hfil$\scriptstyle+$\hfil\cr
					\noalign{\kern-.3ex}
					$\scriptscriptstyle({-})$\cr}%
}}}}

\usepackage{todonotes}

\title{Fluctuating growth rates link turnover and unevenness in species-rich communities}

\shorttitle{Fluctuating growth rates link turnover and unevenness}
\author[1,*]{Emil Mallmin}
\author[1]{Arne Traulsen}
\author[1,2]{Silvia De Monte}
\affiliation[1]{Max Planck Institute for Evolutionary Biology, Plön, Germany}
\affiliation[2]{Institut de Biologie de l'ENS (IBENS), D\'epartement de Biologie, Ecole Normale Sup\'erieure,\newline\phantom{\textsuperscript{2}} CNRS, INSERM, Universit\'e PSL, 75005 Paris, France}
\affiliation[*]{\texttt{mallmin@evolbio.mpg.de}}
\version{preprint v2}
\headerupperright{Mallmin, Traulsen, De Monte (2025)}

\abstract{The maintenance of diversity, the `commonness of rarity', and compositional turnover are ubiquitous features of species-rich communities. Through a minimal model, we consider how these features reflect the interplay between environmental stochasticity, intra- and interspecific competition, and dispersal. We show that, even if species have the same time-average fitness, fluctuations tend to drive the community towards ever-growing unevenness and species extinctions, but self-limitation and/or dispersal allow species-rich states to be sustained. Species abundance--distributions vary systematically in a Buffering--Stabilization parameter plane that describes the relative strength of the underlying ecological processes, and cover different empirically relevant power-law and unimodal shapes. A model describing the effective dynamics of a focal species relates static abundance distributions with turnover dynamics, also when species have different mean fitness. The model suggests how community statistics and time series of individual species can inform on the relative importance of the ecological processes that structure diversity.}

\renewcommand{\sref}[2]{#2}

\begin{document}

\maketitle

Species-rich communities---for instance tropical trees, birds, plankton and microbiomes---pose many fundamental ecological questions. In light of the competitive exclusion principle \cite{Hardin1960,Hutchinson1961,Levin1970}, how do so many species coexist in environments that, seemingly, offer few axes of niche differentiation? At the same time, coexistence is not equitable: at any given time and place of sampling, the vast majority of species have low abundance in comparison to a few highly abundant species \cite{Preston1948,Enquist2019}.
Species abundance distributions (SADs) quantify the spectrum from dominance to rarity in a community, and universally follow a `hollow curve' shape with high index of unevenness \cite{McGill2011}. 
Yet, underlying the relative regularity of SADs is high variability in species' abundances across local samples, and thus in the composition of local communities, for reasons that remains largely unexplained by readily measured environmental factors \cite{Soininen2014,SerGiacomi2018,Mutshinda2016,Rogers2023}.

The precise mechanisms promoting (i) the maintenance of diversity, (ii) the `commonness of rarity', and (iii) pervasive temporal turnover in community composition likely differ between taxonomically and environmentally distant communities. The reality of an ocean-drifting phytoplankter, extracting what amount of light and nutrients it can before being eaten by a zooplankter, is very different from the tropical tree seedling growing to fill the gap in the canopy opened by a full-grown rival struck down by lightning \cite{Smetacek2012}. Still, competition, dispersal, and demographic and environmental forms of stochasticity are general processes that affect all species-rich horizontal communities to varying extents \cite{Vellend2016,McGill2018}. The ubiquity of many biodiversity patterns suggests that an explanation is rightly sought in terms of the interactions of a few such high-level processes. 

Classical and recent theoretical developments \cite{Chesson2000,Barabas2018,West2022,MacArthur1967,Hubbell2001,} point towards the maintenance of diversity as resulting from equalizing processes, making species more similar in competitive ability (e.g.\ via trade-offs in environmental tolerances); stabilizing processes, giving a per-capita competitive advantage to species when rare (e.g.\ less susceptibility to species-specific disease); and a balancing of (local) extinctions with (re-)immigration or speciation. For example, deterministic models of complex communities (notably generalized Lotka-Volterra \cite{Bunin2017,Barbier2018} and consumer-resource models \cite{Advani2018,Cui2021}) show that, under quite general assumptions, species-rich and stable coexistence equilibria are possible if interactions between species are unstructured and `disordered' (equalizing) and
sufficiently weak compared to intraspecific competition (stabilizing). Alternatively, neutral theory \cite{Hubbell2001,HUbbell2005,Volkov2003,Vallade2003} assumes ecological equivalence of species (perfect equalization) and weak or no density-dependence (no stabilization), leaving demographic stochasticity due to the discreteness of individual birth, death, and dispersal events to drive extinctions that are ultimately balanced by metacommunity processes. 

Regarding the commonness of rarity, the stable coexistence regime of disordered competition models produces unrealistically \textit{even} SADs, unless carrying capacities are drawn from an \textit{ad hoc} uneven distribution \cite{Barbier2017x}. In contrast, neutral theory has been celebrated for producing realistic SADs where the shape parameter is given an ecological interpretation as dispersal limitation. But to reliably disambiguate between alternative SAD functional forms in empirical data, e.g.\ lognormal or logseries or a host of alternatives, has proven difficult, and was criticized as a weak test of underlying theories \cite{McGill2003,McGill2007}. Nonetheless, a recent comprehensive analysis \cite{Gao2025} finds a Poisson-sampled power-law with exponential decay at high abundance (`power bend' \cite{Pueyo2006}) to be valid across animal, plant, and microbial communities. The value of the (negative) exponent of the power law section---typically near one for animals and plant communities, and with a median of 1.6 for microbial communities---possibly contains information about underlying processes. For instance, neutral theory predicts an exponent of one, unless generalized to include density-dependent effects \cite{SerGiacomi2018}; logistic growth of independent species with environmental stochasticity \cite{Engen1996a}, or of interacting species with fast stochastic variation in interaction strength \cite{Suweis2024}, results in a gamma distribution, with exponent strictly less than one, unless species are strongly heterogeneous in their demographic rates \cite{Grilli2020,Descheemaeker2021}.

Rapid turnover in community composition can be expected to relate to environmental stochasticity---fluctuations in growth rates due to unmodelled variability in resources, abiotic conditions, predation pressure, etc.---rather than demographic stochasticity alone \cite{Nee2005,Kessler2015b}. Indeed,  environmental but not demographic stochasticity is consistent with the statistics and timescales of empirical abundance fluctuations \cite{Lande2003,Grilli2020}. Time-averaged neutral theory (TAN) \cite{Kalyuzhny2015,Danino2018} augments neutral theory with environmental stochasticity, such that species differ in fitness at any given moment in time, but have comparable fitness averaged over a sufficiently long time. Interestingly, disordered competition models, in an unstable interaction regime leading to deterministic chaos, display abundance fluctuations similar to the effect of environmental stochasticity, but also require the buffering effects of a metacommunity to sustain such dynamics \cite{Roy2020,Pearce2020,Dalmedigos2020,Mallmin2024a,dePirey2024,Blumenthal2024}. The slope of the power-law section of the SAD then depends on the immigration rate, and can be larger than one.

Guided by the preceding insights, we investigate what minimal combination of ecological processes might simultaneously account for (i), (ii), and (iii). A time-averaged neutral model formulated in a Lotka-Volterra framework provides the starting point \cite{Malcai2002,Melbinger2015,vanNes2024,Kessler2024x}. Recently, van Nes et al \cite{vanNes2024} employed such a model to suggest an explanation for the (hyper-)dominance  of a few species in a wide range of community data sets. They draw attention to the `stickiness' effect (called `diffusive trapping' in prior work by Dean and Shnerb \cite{Dean2020}), whereby the scaling of abundance fluctuations biases species that become rare to remain rare. Through an exact mapping to replicator dynamics and to condensation phase transitions in physics, we explain why fluctuating growth rates in fact drive the community rapidly toward unevenness, and eventually monodominance, unless compensated by other processes. We demonstrate how immigration (as in parallel work by Kessler and Shnerb \cite{Kessler2024x}) and self-regulation allow diversity to be maintained long-term. Then, we derive the SAD, which is a generalized inverse Gaussian distribution, interpolating between several empirically relevant cases, including power bend \cite{Jorgensen1982,Sichel1997}. The observed shape varies systematically with two non-dimensional parameters that we interpret as the effective amount of Buffering and Stabilization, respectively. Finally, we relax the TAN assumption to find that moderate heterogeneity in species intrinsic growth rates can lead to large differences in the fluctuation statistics of species. We discuss the empirical relevance and generality of our results.

\vfill

\section*{MODEL}

\section*{Community dynamics}
We consider a pool of $S$ species that in a local community of interest have abundances $n_i(t)$ ($i = 1,2,\ldots, S)$ at time $t$. The net growth rate of a species depends on interactions within the community, and the effect of the broader, time-varying environment; we assume these aspects combine additively. For brevity, we refer to the environmentally determined, density-independent part of the growth rate as \textit{fitness}\footnote{More appropriately \textit{environmental} or \textit{intrinsic fitness}, as distinct from the long-term invasion growth rate notion of fitness in modern coexistence theory.}, denoted by $r_i(t)$. Following classical hypotheses, we take interactions to be dominated by competition (direct or apparent), such that heterospecifics compete with strength $\mu$, and conspecifics with strength $\mu + \varepsilon$. We refer to the special case $\varepsilon = 0$ as \textit{uniform competition}, and call $\varepsilon$ the \textit{excess self-regulation}. Moreover, we consider a small, constant rate of net immigration $\lambda$. Denoting the total abundance by $N(t) = \sum_{j=1}^S n_j(t)$, the above  assumptions define the growth equation
\begin{equation}\label{eq:dotni}
    \dot{n}_i(t) = n_i(t)[r_i(t) - \mu N(t) - \varepsilon n_i(t)] + \lambda.
\end{equation}
We will also consider a spatially explicit metacommunity version of the model, where $M$ patches of local communities are connected through dispersal at rates $d_{\alpha\beta}$ from patch $\beta$ to $\alpha$. For the abundance dynamics of species $i$ in patch $\alpha$, we then replace $\lambda$ above by the species' immigration into the patch minus the emmigration to all other patches,  
\begin{equation}
     \sum_{\beta=1}^M(d_{\alpha\beta} n_{i,\beta}(t) - d_{\beta\alpha} n_{i,\alpha}(t)).
\end{equation}

While the community dynamics encompasses some of the most broadly relevant processes, there are also notable omissions. We do not include demographic stochasticity, but to nonetheless allow for the extinction of rare species we introduce a threshold $n_\text{ext}$ below which abundances are set to zero. Furthermore, species coexistence through the storage effect (i.e.\ noise-induced stabilization) \cite{Chesson1981,Johnson2022} has been precluded, since the fluctuating fitnesses and competition appear additively in the growth rate. For perspectives on these effects in species-rich communities, we refer to several recent works \cite{Danino2018,Pande2022,Kessler2024x}.

\section*{Fluctuating fitnesses}

The fluctuating fitnesses $r_i(t)$ represent the net effect of a complex and time-varying environment that we do not model explicitly. 
For simplicity, we assume the  $r_i(t)$s to be statistically independent between species, and density independent. 
We take each $r_i(t)$ as a coloured noise with expected value $r_i^*$, variance $\sigma_r^2$, and autocorrelation time $\tau$. 
Unless otherwise indicated, we will assume species are \textit{time-average neutral}, meaning $r_i^* = r^*$ \cite{Kalyuzhny2015}. (Fitness variance and autocorrelation will be species-independent throughout.) We let the fitness dynamics follow an Ornstein-Uhlenbeck process

\begin{equation}\label{eq:dotri}
    \tau\dot{r}_i(t) = -(r_i(t) - r_i^*) + \sqrt{2\sigma_r^2 \tau} \dot{W}_i(t),
\end{equation}
where $\dot{W}_i(t)$ formally represents white noise. At stationarity, fitnesses follow a normal distribution $\mathcal{N}(r^*_i, \sigma_r^2)$.

Later, we will identify the parameter combination
\begin{equation}\label{eq:gamma}
\gamma := 2\sigma_r^2 \tau
\end{equation}
appearing in \eqref{eq:dotri} as the \textit{rate of stochastic exclusion}. We will therefore often specify the noise in terms of $(r^*, \gamma, \tau)$ (which implies the value of $\sigma_r$ through \eqref{eq:gamma}). In the \textit{fast environment limit} of $\tau\to 0$, $\sigma_r\to\infty$, while keeping $\gamma$ constant, one obtains a white noise $r_i(t) = r_i^* + \sqrt{\gamma} \dot{W}_i$ (in the Stratonovich stochastic calculus \cite{Pesce2013}). 
Sticking to coloured noise has several advantages, however: we do not implicitly assume environmental fluctuation timescales are fast (perhaps reasonable for elephants, but less so for \textit{E.\ coli}); $\sigma_r$ and $ \tau$ have a clearer biological interpretation than the noise amplitude $\gamma$; and we can ignore subtleties of stochastic calculus convention.

\section*{Model parameters and simulations}
In the fully deterministic, neutral case ($\varepsilon=0$, $\gamma=0$, $\lambda=0$), the total abundance equilibrates at the \textit{carrying capacity} $K := r^*/\mu$. By rescaling abundances, we can set $K=1$. We measure time in units of $1/r^*$, approximately equal to one generation time, which we set to 1 day for ease of communication and without loss of generality.

All model variables and parameters are summarized in \supptabref{tab:symb}{1}; parameter values are specified in the figure captions. The numerical implementation of the model is described in \autoref{sec:num}.

\section*{RESULTS}

\section*{Randomly fluctuating fitnesses drive diversity loss}
To establish a baseline for the effect of fitness fluctuations on species coexistence and diversity, we consider the special case of \eqref{eq:dotni} with uniform competition ($\varepsilon = 0$) and no immigration $(\lambda=0)$:
\begin{equation}\label{eq:neplicator}
    \dot{n}_i(t) = n_i(t)[r_i(t) - \mu N(t)].\\
\end{equation}
Below, we study the dynamics of this system, first in simulation and then analytically, with the following main conclusions.

Coexistence in \eqref{eq:neplicator} is only transient: communities progress toward pronounced unevenness, and eventually monodominance. This is true even in the absence of an extinction cutoff, in which case it takes progressively longer for the identity of the dominant species to change. The stickiness effect forms part of the explanation \cite{Danino2018,vanNes2024}: Because the magnitude of abundance fluctuations is proportional to the current abundance, the rarer a species, the larger (and hence more infrequent) the fitness fluctuation needed to escape rarity. The other part can be traced to the growing variance of fitnesses integrated over time, despite the convergence of time-averaged fitnesses toward $r^*$.

A key measure of the effectiveness of stochastic exclusion is the time $t_c$ it takes for an initially even community to become composed of a few dominant species. We show that it scales as $\ln(S)/\gamma$, with $\gamma$ as in \eqref{eq:gamma}. Whether $t_c$ is a long or short time on the scale of generations depends primarily on $\gamma/r^*$; it is long if relative fluctuations are small ($\sigma_r/r^* \ll 1)$, or if environmental changes are fast compared to generation time ($\tau \ll 1/r^*)$. Remarkably, a community of $S={}$10'000 species would only need twice the time to reach few-species dominance as a 100-species community, all else being equal. Because $t_c \sim \gamma^{-1}$, we will refer to $\gamma$ as the \textit{rate of (stochastic) exclusion}.

\subsection*{Numerical simulations reveal transient diversity }

\begin{figure}[t!]
    \centering
    \includegraphics[width=\columnwidth]{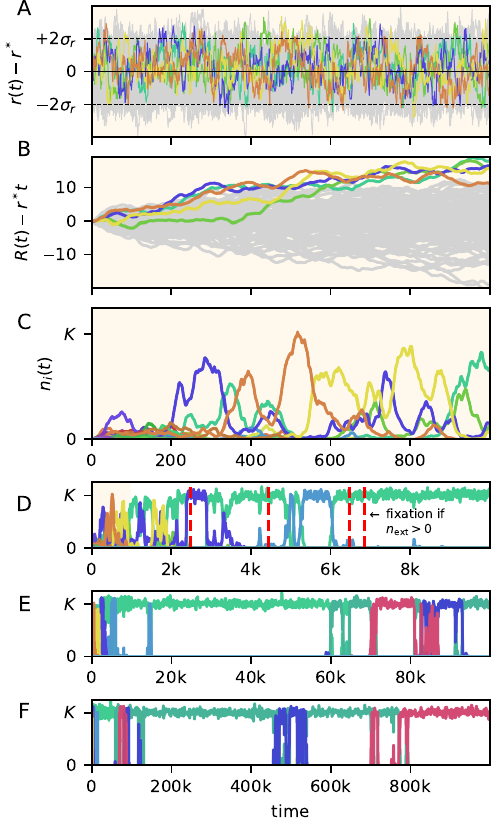}
    \captionof{figure}{\textbf{Simulated community dynamics showing progressive unevenness.} Single simulation run of \eqref{eq:dotni} \& (\ref{eq:dotri}), starting from a perfectly even community. Instantaneous fitness (\textbf{A}) and time-averaged one (\textbf{B}), highlighted in colour for a few highly successful species. The same species are highlighted in the abundance time series (\textbf{C}-\textbf{F}), displayed in increasingly longer time windows. The shown simulation trajectories were generated without an extinction threshold, but the vertical red lines in \textbf{D} indicate when, for the same fitness dynamics as shown, a single species would fixate under different extinction thresholds ($n_\text{ext}/K = 10^{-3}$, $10^{-6}$, $10^{-9}$, $10^{-12}$). Despite a transient with high species diversity, monodominance is readily attained and is fixed for any positive extinction threshold. Simulation parameters are $S=100$, $K=1$, $r^*=1$, $\gamma=0.05$ $\tau=10$.}
    \label{fig:excl}
\end{figure}

To provide intuition on the ecological dynamics of an initially maximally diverse community ($n_i(0) = K/S$, $r_i(0) = r^*$), we simulate \eqref{eq:neplicator} numerically (\autoref{fig:excl}). We observe that, within a few hundred days, a handful of high-abundance species stand out (\autoref{fig:excl}C). While it is difficult to judge any species' success by its instantaneous fitness (\autoref{fig:excl}A), the dominant species can be recognized as having the highest time-integrated fitness since the initial time (\autoref{fig:excl}B). After a few thousand days, the community is dominated by a single species (\autoref{fig:excl}D).
As we observe the abundances over long timescales---from years (\autoref{fig:excl}C), to decades (1D), to centuries (1E), to millennia (1F)---the intervals between exchanges of dominance tend to lengthen. 
Correspondingly, species that are not dominant become increasingly rare, so that, for any positive extinction threshold $n_\text{ext}$, the number of extant species progressively decays until only one species remains (varying $n_\text{ext}/K$ from $10^{-3}$ to $10^{-12}$ has less than an order of magnitude effect on the timescale of fixation; see \autoref{fig:excl}D).
The last surviving species is at no practical risk of stochastic extinction, although, technically, it will vanish eventually.

\begin{figure}[t]
    \centering
    \includegraphics[width=\columnwidth]{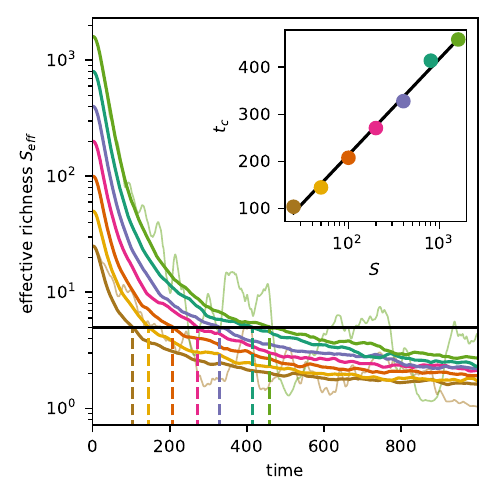}
    \captionof{figure}{\textbf{Decay of the effective species richness} $S_\text{eff}(t)$ (\eqref{eq:seff}), starting from initially even community of 25 to 1600 species. Thick lines show averages over 200 simulations, with the two thin lines illustrating representative single runs for $S=25$ and 1600. The inset shows the time $t_c$ at which the ensemble-averaged  $S_\text{eff}(t)$  has decayed to 5 species, plotted against the initial richness. The initially even community loses its diversity on a timescale of $\ln S$, in agreement with our calculations. }
    \label{fig:exclusion_times}
\end{figure}

We focus next on the timescale for the community to become highly uneven. As a proxy for the number of dominant species, we measure the \textit{effective richness} by Simpson's reciprocal diversity index:
\begin{equation}\label{eq:seff}
    S_\text{eff}(t) := \frac{1}{\sum_i p_i^2(t)},
\end{equation}
where $p_i = n_i/N$ denote relative abundances. In the initially even community $S_\text{eff}(0) = S$, while $S_\text{eff} \to 1$ for large times, signifying monodominance. We measure the time $t_c$ at which the effective richness crosses a threshold of a few species. As shown in \autoref{fig:exclusion_times}, distributions readily grow uneven also in very large communities, with $t_c$ scaling as $\ln S$.

\begin{figure}[t]
    \centering
    \includegraphics[width=\columnwidth]{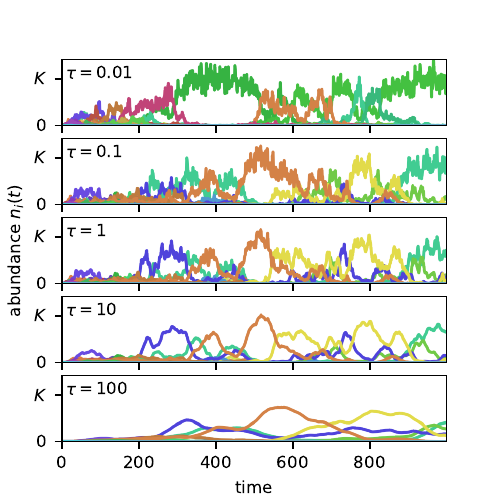}
    \captionof{figure}{\textbf{The main trend in community composition is scarcely affected by the fitness autocorrelation time $\tau$ if the exclusion rate $\gamma$ is fixed.} The random numbers underlying the simulations are identical for all panels. We have fixed $\gamma=0.05$, implying $\sigma_r^2 = 0.125\tau$ by \eqref{eq:gamma}, and then vary $\tau$ between panels.}
    \label{fig:tau}
\end{figure}

The critical time $t_c$ decreases with the rate of stochastic exclusion $\gamma$, as we prove in the next section. Indeed, $\gamma$ essentially sets the `ecological clock' of the model. As shown in \autoref{fig:tau}, when $\gamma$ is fixed, the correlation time $\tau$ alone has little effect on the main dynamical trends, but controls the extent of rapid fluctuations around them. 

\subsection*{A mapping to the replicator equation explains the dynamics of community unevenness}

Despite the large fluctuations of individual species abundances, 
the total abundance $N(t)$ fluctuates only moderately. 
This motivates focussing on the relative abundances $p_i = n_i/N$, which obey the replicator equation \cite{Hofbauer2002} (\autoref{app:replicator})
\begin{equation}\label{eq:peplicator}
    \dot{p}_i(t) = p_i(t) [ r_i(t) - \rho(t)],
\end{equation}
where the community-average fitness $\rho(t) := \sum_j r_j(t) p_j(t)$. This result follows from \eqref{eq:dotni} regardless of the functional form of $r_i(t)$, and is in fact independent of the uniform competition term $\mu N(t)$ or its generalization to any function that has the same value for all species. On the other hand, the strength of the uniform competition $\mu$ constrains the total abundance, whose dynamics
\begin{equation}\label{eq:Ndot}
    \dot{N}(t) = N(t)[ \rho(t) - \mu N(t)]
\end{equation}
is coupled to community composition only through $\rho(t)$.

Key to understanding the dynamics of species composition are the time-integrated fitnesses 
\begin{equation}
    R_i(t) := \int_0^t \diff t'\, r_i(t'),
\end{equation}
as appreciated from the formal solution to \eqref{eq:peplicator}:
\begin{equation}\label{eq:pi_sol}
    p_i(t) = \frac{p_i(0) e^{R_i(t)}}{Z(t)} ,\quad Z(t) := \sum_{j=1}^S p_j(0)e^{R_j(t)}.
\end{equation}
A species $i$ becomes dominant when the factor $e^{R_i}$ makes up a sizeable fraction of the sum of exponentials, so that the question of dominance is essentially one of extreme value statistics. If the gap between the largest (or largest few) $R_i$ and the rest tends to grow in time, then eventually---and, indeed, rather soon due to the exponentiation---the corresponding species will come to dominate. If species differ in their expected fitnesses, the one with larger average fitness eventually wins deterministically (competitive exclusion). In the time-averaged neutral case with random fitness fluctuations following \eqref{eq:dotri}, after a transient of length comparable to $\tau$, the $R_i$s diverge at rate $\gamma$ (see \eqref{eq:Rk} in \autoref{sec:rstats}), which thus controls the speed at which community unevenness develops. Also, the aging dynamics observed for $n_\text{ext}=0$, where changes in dominance become increasingly rare, is explained by the property of Brownian motions $W_i$ to return to the origin in finite time despite the growing variance. In contrast, there is no asymptotic monodominance if fluctuations are perfectly periodical, e.g. $r_i(t) = r^* + \sqrt{2}\sigma_r \cos(t/\tau - \phi_i)$, because the variance among $R_i$s remains bounded.

We note that \eqref{eq:pi_sol} has the form of the Boltzmann distribution from equilibrium statistical physics. Indeed, in \autoref{sec:PT} we show how the  ecological model with Gaussian fitness fluctuations can be exactly mapped to the `random energy model' of a spin glass, for which many properties have been calculated in the large-system limit \cite{Derrida1980}. The spin glass exhibits a condensation phase transition at a critical temperature, which is mathematically analogous to the community unevenness transition at a critical time \eqref{eq:gtw} scaling as
\begin{equation}\label{eq:tc}
    t_c \sim \frac{\ln S}{\gamma},
\end{equation}
on the assumption that $\tau \ll t_c$. This analytical result matches the scaling of $t_c$ with $\ln S$ that we observed in simulations (\autoref{fig:exclusion_times}). 

The dynamics of community unevenness can also be understood by looking at a single \textit{focal species}. For a community of two species \cite{Danino2018}, using $p_2 = 1 - p_1$ in \eqref{eq:peplicator},
\begin{equation}\label{eq:p1}
    \dot{p}_1(t) = \Delta r_1(t) p_1(t)[1- p_1 (t)],
\end{equation}
with $\Delta r_1(t) := r_1(t) - r_2(t)$, which is independent of any species' abundance. As the relative abundance of the focal species 1 approaches either $0$ or $1$, the dynamics slows down, keeping the species generally closer to these extremes than at any intermediate value. We show in \autoref{sec:stickiness} that \eqref{eq:p1} holds for a focal species also in an $S$-species community, given a generalized form of $\Delta r_1$. Consider the sub-community of all species \textit{except} the focal one, and denote by $\rho_{\backslash 1}(t)$ the mean fitness in this subcommunity (i.e., where relative abundances are normalized only with respect to the $S-1$ non-focal species). Then \eqref{eq:p1} holds for 
\begin{equation}
    \Delta r_1(t) := r_1(t) - \rho_{\backslash 1}(t).
\end{equation}
Unlike the two-species case, $\Delta r_1$ now has a negative bias: $\rho_{\backslash 1}$ is weighted towards the species with higher abundances, which tend to have higher-than-average growth rates. Thus, all species are biased towards rarity, but since relative abundances are normalized---a constraint  enforced through correlations between all the $\Delta r_i$s---some species will buck the trend and seize a large fraction of the total abundance. We note that similar dynamical aging appears in a deterministic model where $\Delta r_1$ encompasses heterogeneous species interactions \cite{dePirey2023}.

\begin{figure*}[t!]
    \centering
    \includegraphics[width=\columnwidth]{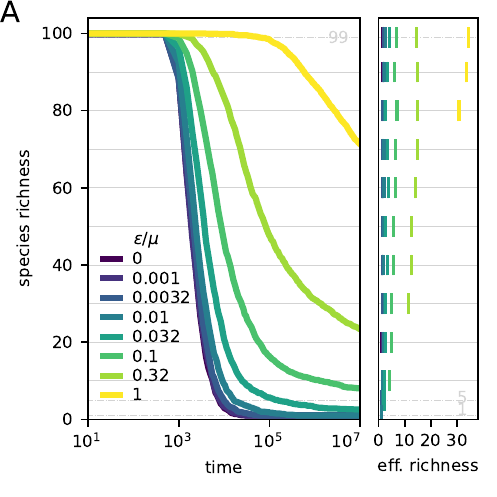}%
    \includegraphics[width=\columnwidth]{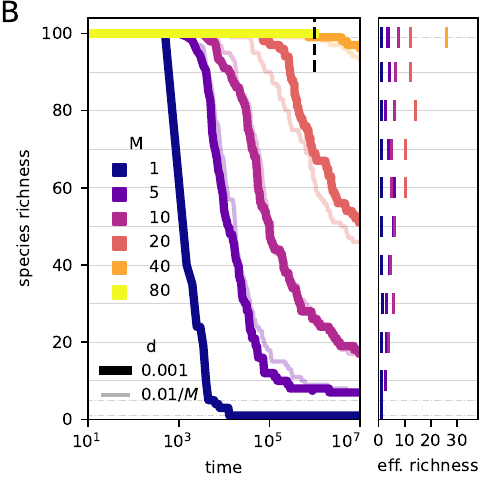}
    \caption{\textbf{Loss off species richness (large panel) and effective richness \eqref{eq:seff} (side panel) over time}. The effective richnesses are plotted for the times at which the absolute abundance crosses the level values indicated by horizontal grey lines.  The average diversity decay over 20 simulation runs is plotted for different values of excess self-regulation $\varepsilon$ (\textbf{A}), and in metapopulations with different number of patches $M$ and migration rates $d$ (\textbf{B}). Other parameters are as given in \autoref{fig:excl} with $n_\text{ext}/K = 10^{-12}$. }\label{fig:coexistence}
\end{figure*}

\section*{Species loss is drastically slowed by intraspecific limitation or metacommunity buffering}
As we have demonstrated, environmental stochasticity can drive `commonness of rarity' and turnover of composition, but only transiently. Long-term maintenance of species richness requires local coexistence mechanisms \cite{Chesson2008,Barabas2018}, or extinction--colonization balance \cite{MacArthur1967,Hubbell2001}. We therefore consider the effects of additional  intraspecific limitation or metacommunity dispersal on diversity.

We suppose intraspecific competition exceeds interspecific competition by an amount $\varepsilon>0$:
\begin{equation}\label{eq:model_suppression}
    \dot{n}_i(t) = n_i(t)[r_i(t) - \mu N(t) - \varepsilon n_i].
\end{equation}
This introduces negative frequency dependence, such that a species is penalized (favoured) when its relative abundance is above (below) $1/S_\text{eff}$ (see \eqref{eq:ptilde}, \autoref{sec:nondim}). In principle, any $\varepsilon>0$ stabilizes coexistence (\autoref{sec:dist}), but only in the absence of an extinction threshold. As we allow for extinctions, increasing $\varepsilon/\mu$ from 0 to 1 increases the timescale of substantial loss of species richness by many orders of magnitude (\autoref{fig:coexistence}A). The effective species richness remains roughly constant until constrained by the absolute richness, as the rare species headed toward extinction have a marginal effect on the rest of the community. It is therefore reasonable to consider the community as quasi-stationary on timescales that can indeed be very long even when self-regulation is weak. Further increasing $\varepsilon/\mu$ to around $3S \sigma_r/r^*$ (= 15 with default simulation parameters) would allow essentially all species to coexist deterministically if the fitnesses were suddenly frozen, i.e.\ drawn statically from the stationary distribution (\autoref{app:fp}); then stochastic exclusion does not occur at all, in practice.   

Alternatively, we introduce a metacommunity buffering effect through self-consistent dispersal among $M$ patches: 
\begin{equation}\label{eq:model_patches}
    \dot{n}_{i,\alpha} = n_{i,\alpha} ( r_{i,\alpha} - \mu N_\alpha) + \sum_{\beta=1}^M(d_{\alpha\beta} n_{i,\beta} - d_{\beta\alpha} n_{i,\alpha}).
\end{equation}
Naturally, the rates of local and regional extinction will depend on the number of patches and the topology of the network, the correlation in environmental conditions between patches, the rates of dispersal, and the extinction threshold; enough factors to make a systematic analysis challenging. In \autoref{fig:coexistence}B we only consider a fully connected patch network, uncorrelated fitnesses, and vary either the net dispersal rate ($d$) or the dispersal per patch $(d/M)$. In either case, every doubling of the number of patches leads to about one more order of magnitude in the time it takes to lose species richness in a given patch. Indeed, related metacommunity models have found species lifetimes to grow exponentially with the number of patches \cite{Roy2020,Lorenzana2024}.

\section*{A modified power-law abundance distribution is maintained by turnover of rare and dominant species}

\begin{figure*}[t!]
    \centering
 
    \includegraphics[width=0.97\textwidth]{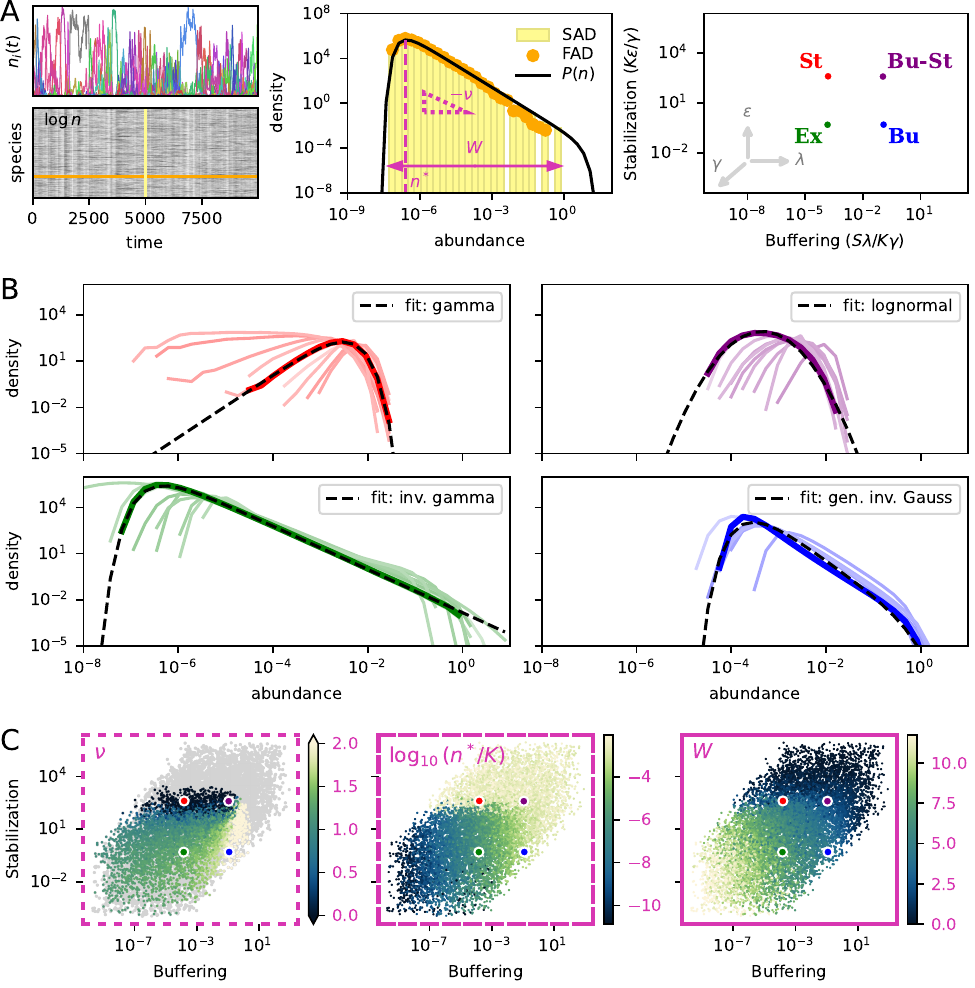}

    \caption{ {\bf Variation in the shape of abundance distributions across simulated communities with varying base parameters}. \textbf{A} Left: An example times series with the corresponding abundance matrix (times $\times$ species). Middle: Different abundance distributions constructed from the abundance matrix: the `snapshot' species--abundance distribution (SAD, yellow histogram); the frequency--abundance distribution (FAD, orange symbols) for one arbitrary species; the predicted stationary distribution \eqref{eq:P(n)} of the focal species model with effective noise statistics measured from the data (solid black line). Three key features of the distributions are highlighted (pink): the number of decades $W$ spanned by the SAD; the modal abundance class $n^*$; and the downward slope of the power-law section, as defined by the formula \eqref{eq:nu}. Right: A reduced parameter space, where each point in the 7-dimensional base parameter space is mapped to a value of Buffering (horizontal axis) and Stabilization (vertical axis). The inset arrows show the direction of movement as the indicated parameter ($\varepsilon$, $\lambda$, or $\gamma$) is changed  while all others are held constant. Four points have been marked as references and named according to the dominant process: Ex (Exclusion), St (Stabilization), Bu (Buffering), Bu-St (Buffering-and-Stabilization). \textbf{B} Four panels corresponding to the four reference points, each showing ten SADs (from the ten simulations whose parameters lie closest to the reference points). One SAD has been highlighted (bold) and fitted with a particular distribution (gamma, lognormal, inverse gamma, or generalized inverse Gaussian). \suppfigref{fig:fad_grid}{2} shows the morphing of one class of shape into another as we move in the Buffering--Stabilization space.  \textbf{C} Variation of SAD features across the Buffering--Stabilization space. Each point represents one simulation. Parameters were sampled (log-)uniformly to vary over orders of magnitude: $S \in [100,1000]$, $\log_{10} \gamma \in [-4,2]$, $\log_{10}\tau \in [-2,2] $,  $\log_{10}\varepsilon \in [-2,2] $, $\log_{10} \lambda \in [-10,-4]$. Units are adapted so that $K=1$ and $r^*=1$. In the $\nu$ panel, points were excluded (gray) if the focal-species SAD has a goodness-of-fit below $85\%$ (see \suppfigref{fig:Pfit}{1}) or the distribution covered less than two decades ($W < 2$). Note that colour scale for the exponent is capped in the range $[0, 2]$, so that all negative values appear in the same color (dark blue).}    
    \label{fig:distributions}
\end{figure*}

Given the radical slowdown of diversity loss achievable by modest amounts of excess self-regulation or dispersal, we consider in the following the single-patch dynamics \eqref{eq:dotni} \& \eqref{eq:dotri} without extinction cut-off, which has a true stationary state. We look at two empirically relevant statistics: the abundance distributions displayed by individual species over long stretches of time (frequency--abundance distribution, FAD), or by all species of the community at a snapshot in time (species--abundance distribution, SAD). Their relation is illustrated in \autoref{fig:distributions}A. While all species fluctuate in abundance over time, the SAD retains its general shape across snapshots, which appears to be a subsampling of the FAD. Moreover, all species have identical FAD if compared for a sufficiently long time, due to species-symmetry of the model parameters. Thus, for large, time-average neutral communities, the FAD and SAD essentially coincide.

Seeking to derive the form of the abundance distribution, we consider the dynamics of a focal species, for which the influence of the rest of the community is treated as part of an `effective' fluctuating environment:
\begin{equation}\label{eq:focal}
    \dot{n} = n(r_\text{eff}(t) - \varepsilon n) + \lambda.
\end{equation}
We take $r_\text{eff}(t)$ to be an Ornstein-Uhlenbeck process like \eqref{eq:dotri}, but with mean $r_\text{eff}$, variance $\sigma_{r_\text{eff}}^2$, and autocorrelation time $\tau_\text{eff}$ (from which we define $\gamma_\text{eff} = 2 \sigma_{r_{\text{eff}}}^2 \tau_\text{eff}$, as before). These statistics are tuned to approximate those of $r_i(t) - \mu N(t)$ (see \autoref{sec:dist}). In the fast-environment limit, the stationary distribution is
\begin{equation}\label{eq:P(n)-lim}
    P(n) \propto n^{-\nu} e^{- n/a - b / n},
\end{equation}
combining a power-law section with exponent
\begin{equation}\label{eq:nu}
    \nu = 1 - \frac{2 r^*_\text{eff}}{\gamma_\text{eff}}
\end{equation}
and downward `bends' beyond sufficiently high or low abundances
\begin{equation}\label{eq:ab}
    a = \frac{ \gamma_\text{eff}}{2 \varepsilon},\quad b = \frac{2 \lambda}{ \gamma_\text{eff}}. 
\end{equation}
In the general case with finite noise correlation the distribution can be solved for approximately  and is also a `bent' power law (\autoref{sec:dist}).

If we consider $\nu$ (real), $a$ $ (> 0)$, and $b$ $ (> 0)$ as independent parameters, then \eqref{eq:P(n)-lim} is known as the {generalized inverse Gaussian} (GIG) distribution \cite{Jorgensen1982}. It contains as special or limiting cases: the inverse Gaussian distribution, the gamma distribution, a continuous interpolation of the {powerbend} distribution, pure power law, and the lognormal distribution. All of these (including the GIG itself \cite{Sichel1997}) have been considered as `underlying' SADs in the ecological literature, commonly mixed with a Poisson distribution to model sampling effort \cite{McGill2007}. The principal differences between many alternative SADs is the presence and location of a mode (which may or may not be detectable in small-size samples) and the extent and slope of a power law section.

For us, the GIG parameters $\nu,a,b$ are \textit{not} independent, however. They depend on the effective parameters $r_\text{eff}$ and $\gamma_\text{eff}$, which are in turn determined implicitly through the complex community dynamics by the seven base parameters $S,r^*,K (=r^*/\mu),\gamma,\tau,\varepsilon,\lambda$. Of the possible shapes that the GIG affords, which are actually realized by the model, and how can we interpret the shape in terms of the underlying ecological processes? One approach to infer the effective parameters from the base set is by imposing a self-consistency relation on the focal-species model (see \autoref{sec:dist} for an explanation and tractable special case). In the following, we instead note that a dimensional analysis of \eqref{eq:dotni} (\autoref{sec:nondim}) indicates that the variation in distribution shape is mainly captured by two composite parameters:
\begin{equation}\label{eq:BS}
    B = \frac{S\lambda}{K\gamma}\quad \text{and}\quad 
    \Sigma = \frac{K\varepsilon}{\gamma}.     
\end{equation}
We interpret the first ($B$) as a \textit{Buffering} dimension: it increases with the net immigration rate  ($S\lambda$). The second ($\Sigma$) we interpret as a \textit{Stabilization} dimension: it increases with the excess self-regulation ($\varepsilon$). Both are reduced by increasing environmental noise ($\gamma$), which reflects  the notion that the amount of buffering or stabilization is relative to the strength of exclusionary processes. Increasing carrying capacity ($K$) lowers buffering, as the relative abundance increase from an immigrating individual is proportionally less, and increases stabilization, as more species attain abundances where self-regulation is strong. 

We simulate 10'000 communities with base parameters drawn randomly from a wide range of values, and quantify the distribution shape by three indices (illustrated in \autoref{fig:distributions}A): the width $W$ in number of abundance decades; the modal abundance $n^*$; and the power-law exponent $\nu$ as defined by \eqref{eq:nu}. 
By projecting these indeces in the  Buffering--Stabilization space, we confirm that the compound parameters $B$ and $\Sigma$ capture most of the variation in SAD shape \autoref{fig:distributions}C. 
The shape varies continuously, but four main types can be identified (\autoref{fig:distributions}B), corresponding to different ecological regimes.

\textit{Exclusion regime}. When buffering and stabilization are both small, we observe a power-law section spanning many orders of magnitude---a few species are highly dominant, and the rest are rare. In the complete absence of buffering or stabilization (as in \eqref{eq:neplicator}), fluctuations drive an ever-widening power law of exponent approaching 1 (\suppfigref{fig:sad_spread}{3}; see also the Poisson-process limit law \cite{Eliazar2020}).

\textit{Buffered regime}. Buffering increases the lower bound of abundances, and the power-law section steepens as a larger fraction of species accumulate at the immigration threshold. Here we can find exponents $\nu$ in the upper empirical range.  

\textit{Stabilized regime}. With sufficient stabilization, the mode of the distribution drastically shifts from lying close to the immigration threshold to approaching the single-species carrying capacity (which is simultaneously reduced by self-regulation). This shift coincides with the power-law exponent changing sign. When it is positive, we observe a gamma distribution.

\textit{Buffered-and-stabilized regime}. When buffering and stabilization are both prominent, abundances are tightly constrained between the lower limit due to immigration, and an upper limit set by self-regulation. The resulting shape is well approximated by a lognormal.

Because the SAD here reflects an FAD common to all species, its shape is closely linked to temporal beta diversity. The average Bray-Curtis similarity $\text{BC}(t)$ of communities observed a time $t$ apart (a quantity that has been applied to ecological time series \cite{Fuhrman2015}), decays from 1 towards a limiting value that depends on the shape of the distribution, thus chiefly on the position in the Buffering--Stabilization plane (\suppfigref{fig:bc_grid}{4}). Indeed, comparing the community composition at two time points sufficiently far apart amounts to randomizing the species ranks of the second sample with respect to the first, while preserving the SAD. For the power-law shape in the exclusion-dominated region, the asymptotic similarity is smaller than when abundances are more narrowly distributed around their mode ($< 0.15$ compared to $>0.5$). The \textit{rate} at which BC similarity decays, however, depends on the actual value of the exclusion rate $\gamma$. For large values, specifically, the dominance of exclusion is associated to a faster species turnover and larger abundance fluctuations.

\begin{figure*}[t]
    \centering
    \includegraphics[]{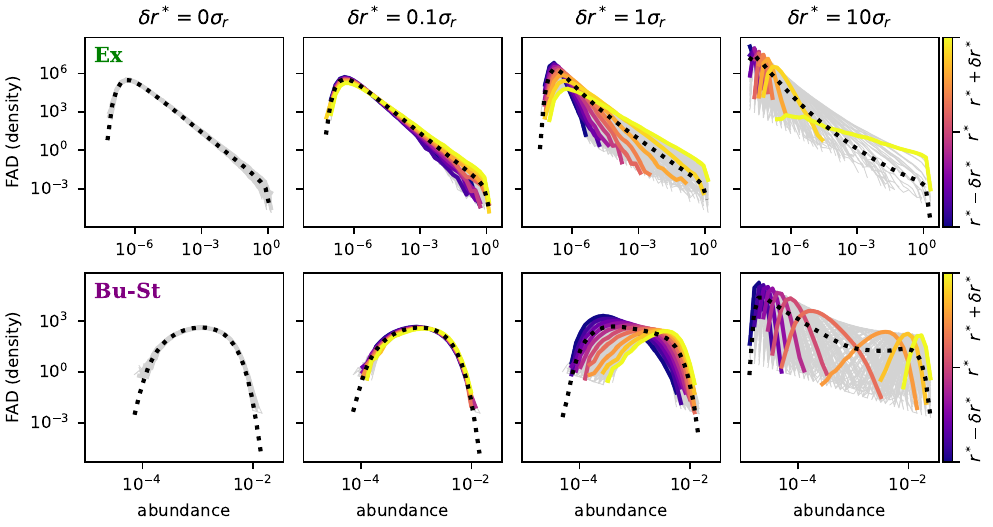}
    \caption{\textbf{Emergence of frequent and infrequent species under breaking of time-averaged neutrality.} The two rows report the same numerical protocol but starting from different sets of model parameters, corresponding to the Exclusion regime and Buffered-and-Stabilized regime (points Ex and Bu-St in \autoref{fig:distributions}). Time-averaged neutrality is broken by drawing the $r_i^*$s uniformly at random from the interval $[r^*-\delta r^*, r^* + \delta r^*]$, with $\delta r^*$ varying by column. In each panel, the frequency--abundance distribution (FAD) of each individual species is plotted, with a subset of species shown in colour according to their $r^*_i$; the dotted dashed lines are the species-averaged FADs (equal to the time-averaged SAD). Simulation parameters used are $S=500$, $r^*=1$, $K=1$, $\gamma = 0.05$, $\tau=10$, and for Ex, Bu-St, respectively, $\varepsilon=0.05, 50$; $\lambda=10^{-8},3.2\times 10^{-5}$. Simulations were run for 500'000 time units.}
    \label{fig:nontan}
\end{figure*}

\section*{Heterogeneous fitnesses create persistent biases in species rarity}
Finally, we relax the assumption of time-averaged neutrality in order to test the robustness of our results when species within the same community differ in their long-term abundance statistics, as expected in real communities. 

We take two reference parameter sets corresponding to the Exclusion and Buffered-and-Stabilized regimes (same as in \autoref{fig:distributions}). For each case, we draw the species-specific fitness averages $r_i^*$ from a uniform distribution in the range $r^* \pm \delta r^*$, but keep all other parameters identical for all species. The fitness distributions are now same-width normal distributions with different means; how much they overlap is controlled by $\delta r^* / \sigma_r$, which we vary from $0$ (total overlap) to $10$ (small overlap). As this ratio increases, the species-specific FADs separate, and species of lower (higher) $r_i^*$ become biased towards rarity (dominance); see \autoref{fig:nontan}. For the heavy-tailed distribution in the Exclusion regime, the time-averaged SAD (coinciding by definition with the species-averaged FAD) changes little during this splitting. In the Buffered-and-stabilized regime, the initially lognormal time-average SAD widens to develop a power-law trend. The breaking of TAN also leads to a smaller turnover as measured by the BC index limit, as species become more constrained in their fluctuations, whether biased towards rarity or dominance (\suppfigref{fig:bc6}{5}).

Focussing on the emerging 
differences, we note that a spread in mean fitness within the limits of a doubling/halving of the average (the rightmost panels  of \autoref{fig:nontan} have $r_i^* \in [0.5,1.5]$) is able to produce a distribution of species mean abundances spanning several orders of magnitude (\suppfigref{fig:mean_dist}{6}). The shape of the FADs can also differ between species. As FADs split, the `frequent' species' FADs are more gamma-like, whereas `infrequent' ones have a longer right tail, as particularly evident in the St-Bu example. The possible FAD shapes of different species are still described by the focal-species model \eqref{eq:focal}, but now the species have different $r_\text{eff}^*$, set by $r_i^* - \mu \overline{N}$. Thus, based on its expected fitness, a species may be relatively more or less constrained by immigration or self-regulation. 

\section*{DISCUSSION}

We have sought to understand how environmental stochasticity, intra- and interspecific competition, and dispersal relate to three features of natural communities: the coexistence of many species, the commonness of rarity (reflected in heavy-tailed SADs), and species turnover.

Even under the equalizing assumptions of uniform competition and time-averaged neutrality, environmental stochasticity drives the community towards ever-greater unevenness (and species extinctions), unless countered by other processes. Contrary to strictly neutral dynamics driven by demographic noise \cite{Kessler2015b}, the timescale on which a few species rise to dominance is independent of total carrying capacity \cite{Dean2020}; its determinants are the variance of fitness fluctuations and their correlation time, beside a weak dependence on species richness. The stochastic exclusion dynamics reflects extreme-value statistics that also underlie phase transitions in disordered systems in physics. The model thus reinforces the well-subscribed notion that particular coexistence mechanisms are needed to explain natural diversity, at least in communities where competition is pervasive. We observed how increasing intraspecific suppression, or dispersal within a metacommunity (promoting a spatiotemporal buffering effect \cite{Loreau2003,Lorenzana2024}) allows both absolute and effective species richness to be preserved on super-generational timescales.  

We then formulated a simple focal-species model as a means to connect the fluctuation statistics of individual species with community statistics, especially the SAD. Under time-average neutrality, it is a `bent power law', close to the three-parameter generalized inverse Gaussian distribution (GIG) postulated as a flexible SAD model over three decades ago \cite{Sichel1997}. Emerging from our dynamical model, its parameters are not independent: instead we find that two compound parameters largely account for the variability in distribution shape. We interpret them as measures of \textit{Buffering} (proportional to immigration rate) and \textit{Stabilization} (proportional to excess self-regulation)---similar in spirit to dispersal-limitation and niche--neutral axes \cite{Gravel2006,Fisher2014,Haegeman2011,Leibovich2022}, but measured with respect to strength of environmental fluctuations as opposed to neutral drift. The SAD approaches a wide power law of exponent $\nu = 1$ when both buffering and stabilization are weak; stabilization promotes a gamma distribution shape ($\nu < 1$, typically negative and producing a non-zero mode); buffering makes the power law section narrower and steeper ($\nu > 1$); and the combination of both produces a peaked distribution well-fitted by a lognormal. The focal-species model lends an ecological interpretation to the exponent. It will be close to unity when mean effective fitness (encompassing its intrinsic growth rate and community interactions) is close to zero, or if effective environmental stochasticity is very large. Exponents $\nu > 1$, as for plankton, require \textit{negative} mean effective fitness. In that case, species richness can only be sustained by immigration, which is indeed a major process of planktonic communities. 

In reality, some species are persistently more common than others \cite{Saether2013}. 
For example, an estuarine fish community contained both established `core' species, and `occasional', non-establishing immigrants \cite{Magurran2003}.
 When we relaxed the assumption of time-averaged neutrality by allowing some dispersion in species' expected fitness, we could produce order of magnitude differences in species mean abundance, and qualitatively different shapes of their frequency--abundance distributions (FADs), with implications for the SAD that remain to be systematically explored. The amplification of moderate difference in species demographic parameters into stark differences in their abundance patterns due to within-community feedbacks speaks for the difficulty in predicting species-level composition of natural communities. Other relevant sources of heterogeneity, e.g.\ differences in intraspecific \cite{Yenni2017,Rovere2019} or interspecific interactions \cite{May1972,Buche2022,Mallmin2024a,Kessler2024x}, further add to this complexity.

As a highly aggregated measure, the SAD may provide limited information for disambiguating between alternative theories \cite{McGill2007}. A per-species spectrum of FADs provides a richer picture \cite{Grilli2020}, but requires highly resolved time-series and does not give direct information about community timescales. Complementing the abundance distributions, we studied the decay in community similarity over time as an empirically relevant measure of turnover \cite{Kampichler2012,Fuhrman2015}. In steady state, the long-time similarity to an initial composition is higher in communities with a narrower SAD, and where species are more constrained in their range of fluctuations due to long-term differences in fitness. The decay rate is closely connected to the strength of environmental fluctuations. Investigating empirically if there are correlations between SAD shape parameters and turnover measures would be a worthwhile extension of recent macroecological surveys \cite{Blowes2019,Pinsky2025,Gao2025}.

Ultimately, comparing empirical community data and theoretical predictions over a whole suite of simultaneous patterns (abundance distributions, turnover, extinction times, species correlations, \ldots) will elucidate community assembly. Such an endeavor is well underway for microbial communities \cite{Ji2020,Grilli2020,Descheemaeker2020,Descheemaeker2021,Sireci2023,Wang2023,Maskawa2025}. Models have varied in the emphasis on species interactions \cite{Wang2023}, parameter heterogeneity \cite{Grilli2020}, or spatiotemporal coarse-graining \cite{Maskawa2025}, but generally encompass fluctuating growth rates. Despite the differences in ecological mechanism, stochastic variation in intrinsic fitness, in interaction rates, or interaction-driven chaos, lead to more or less identical forms of the focal species dynamics  \cite{Mallmin2024a,dePirey2024,Suweis2024}. This `multiple realizability' of fluctuating growth rates points towards the generality of our results.

\vfill

\section*{ACKNOWLEDGEMENTS}
The authors thank Nadav Shnerb and David Kessler for insightful discussions and sharing \cite{Kessler2024x} ahead of publication.

\appendix

\section{Statistics of fitness fluctuations}\label{sec:rstats}

Integrating the Ornstein-Uhlenbeck process, \eqref{eq:dotri}, from time $t_{k}$ to $t_{k+1} = t_k + \Delta t$ (dropping the index $i$ for brevity),
\begin{equation}\label{eq:rk}
    r(t_{k+1}) =  r(t_k)e^{-\Delta t/\tau} + r^* (1- e^{-\Delta t/\tau}) + \sigma_r 
    u(\Delta t/\tau) X_k,
\end{equation}
where $X_k \sim \mathcal{N}(0,1)$ and $u(s) = \sqrt{ 1 - e^{-2s}}$. The integrated fitness satisfies \cite{Gillespie1996} 
\begin{multline}\label{eq:Rk}
    R(t_{k+1}) =  R(t_k) + r^*\Delta t + \tau (r(t_k) - r^*) (1 - e^{-\Delta t/\tau}) \\
     +  \sqrt{\gamma \Delta t} v(\Delta t/\tau) Y_k,
\end{multline}
where $Y_k \sim \mathcal{N}(0,1)$ and 
\begin{equation}\label{eq:v}
    v(s) = \sqrt{1 - \frac{1}{2 s}\left( 3 + e^{-2s} - 4 e^{-s} \right)} \sim \begin{cases}
        \frac{1}{\sqrt{3}}s & s \ll 1\\
        1 & s \gg 1
    \end{cases}.
\end{equation}
Note that $X_k, Y_k$ have a correlation $C(\Delta t / \tau )$ with
\begin{equation}\label{eq:C}
    C(s) = \frac{1 - 2e^{-s} + e^{-2s}}{\sqrt{\left( 1 - e^{-2s} \right)\left( 2s - \left( 3 + e^{-2s} - 4 e^{-s} \right) \right)}},
\end{equation}
which is also the correlation between $R(\Delta t)$ and $r(\Delta t)$. It takes $t \approx 50\tau$ for $C(t/\tau)$ to reduce from $C(0)=\sqrt{3}/2$ to a 10\% correlation, and $5000\tau$ to a 1\% correlation. 

\section{Numerical integration}\label{sec:num}
The numerical integration of \eqref{eq:dotni} was done with the scheme
\begin{equation}
    n_{i}[t+\Delta t] \leftarrow n_i[t] \exp\big\{ \Delta R_i[t] - \Delta t( \mu N[t] + \varepsilon n_i[t]) \big\} + \lambda \Delta t.
\end{equation}
It ensures positivity of abundances, and, if $\varepsilon=0$ and $\lambda=0$,   reproduces the analytically exact path probabilities for relative abundances regardless of step size $\Delta t$. A step-size of $\Delta t = 0.01$ was used throughout. For Ornstein-Uhlenbeck noise, $\Delta R_i$ can be sampled from \eqref{eq:Rk} in tandem with \eqref{eq:rk}, involving no approximations.

In the metacommunity version of the model, \eqref{eq:model_patches},  
\begin{multline}
    n_{i\alpha}[t+\Delta t]  {}\leftarrow \left( \sum_{\beta=1}^M n_{i\beta}[t] \mathcal{D}_{\beta \alpha} \right) \times \\
    \exp\left\{ \Delta R_{i\alpha}[t] - \Delta t\left(\mu N_\alpha[t] + \varepsilon n_{i\alpha}[t]\right) \right\},
\end{multline}
where
\begin{equation}
    \mathcal{D}_{\alpha\alpha} = 1 - e^{-d_\alpha \Delta t},\quad \mathcal{D}_{\alpha\beta}= \frac{d_{\beta\alpha}}{d_\alpha}e^{-d_\alpha \Delta t}\quad (\beta \neq \alpha).
\end{equation}

\section{Derivation of the replicator equation}\label{app:replicator}
To derive \eqref{eq:peplicator}--(\ref{eq:Ndot}) from \eqref{eq:dotni}, consider more generally
\begin{equation}
    \dot{n}_i(t) = n_i(t)[r_i(t) - g(\vec{n}(t), \vec{r}(t), t)].
\end{equation}
By summing over $i$ we get immediately
\begin{equation}
    \dot{N}(t) = N(t)[ \rho(t) - g(\vec{n}(t), \vec{r}(t), t)],
\end{equation}
with $\rho = \sum_j r_j n_j/N$. Applying the chain rule to $p_i = n_i/N$,
\begin{equation}
    \dot{p}_i = \frac{\dot{n}_i}{N} - p_i \frac{\dot{N}}{N} =  \frac{n_i}{N}(r_i - g) - p_i (\rho - g) = p_i(r_i - \rho).
\end{equation}
The cancellation of $g$ hinges upon it being identical for all species (i.e.\ neutral).

\section{Derivation of focal-species replicator equation}\label{sec:stickiness}
Here we derive the focal-species \eqref{eq:p1} revealing the stickiness effect. With $p_i$, $i=1,\ldots,S$, the relative abundances in the full community, define the relative abundance $p_{i\setminus 1}$, $i=2,\ldots,S$, in the sub-community excluding species $1$:
\begin{equation}\label{eq:pisub}
    p_{i\setminus 1} := \frac{n_i}{\sum_{j=2}^S n_j} = \frac{p_i}{1 - p_1},\quad i>1. 
\end{equation}
Note that $p_{i\setminus 1}$ is independent of the fitness of species 1 since
\begin{equation}
    p_{i\setminus 1} = \frac{p_i(0)e^{R_i}}{\sum_{j>1} p_j(0) e^{R_j}}.
\end{equation}
The mean fitness in the full community is
\begin{equation}\label{eq:stick_2}
    \rho = p_1 r_1 + \sum_{i=2}^S p_i r_i = p_1 r_1 + (1-p_1)\rho_{\setminus 1},
\end{equation}
where we have substituted $p_{i\setminus 1}(1-p_1)$ for $p_i$ according to \eqref{eq:pisub}, and identified
\begin{equation}
    \rho_{\setminus 1} := \sum_{i=2}^S  p_{i\setminus 1} r_i.
\end{equation}
Substituting the above expression for $\rho$ in the replicator equation, \eqref{eq:peplicator}, for $i=1$ reproduces \eqref{eq:p1}. 

\section{Unevenness as a phase transition}\label{sec:PT}
Here we give more details on the derivation of the timescale of the uneveness transition, and the mapping to a spin glass. 

At any fixed time $t$, the distribution of $p_i(t)$, \eqref{eq:pi_sol}, has the mathematical form of a Boltzmann distribution, 
\begin{equation}
    p_i^\text{B} = \frac{e^{-\beta E_i}}{Z^\text{B}}.
\end{equation}
The $S$ species correspond to as many configurations of some physical system, for instance the $S=2^J$ possible configuration of $J$ Ising spins, in contact with a heat bath at inverse temperature $\beta$. $\ln n_i(0) + R_i(t)$ is mapped to $\beta E_i$, with $E_i$ the energy of the physical configuration. We assume fitness fluctuations according to \eqref{eq:dotri}; that $n_i(0)$ are log-normally distributed with log-variance $\sigma_{n,0}^2$; and that the $r_i(0)$s are normal with variance $\sigma_{r,0}^2$. Then $\ln n_i(0) + R_i(t)$ is normally distributed at any $t \geq 0$. The assumption of normal energy levels with variance $J/2$ defines the 
random energy model (REM) of a spin glass \cite{Derrida1980,Derrida1981}. The mapping from the ecological model to the REM is fixed by choosing $\beta = \beta(t)$ according to   
\begin{equation}
    \beta^2 J/2 = \Var{n(0) + R(t)},\quad 2^J = S.
\end{equation}

In the `thermodynamic limit' $J \to \infty$, the REM exhibits a condensation phase transition at the critical inverse temperature $\beta_c = 2 \sqrt{\ln 2}$, separating a phase dominated by the energy ground state and one where almost all states are equiprobable. Thus, in the community model there is a dominance transition at the critical time $t_c$ given implicitly by $\beta_c = \beta(t_c)$. From our assumptions, using \eqref{eq:Rk},
\begin{equation}
    \left(\frac{\beta(t)}{\beta_c}\right)^2 = \frac{1}{2\ln S} \left(  \sigma_{n,0}^2 + \tau^2 (1 - e^{-t/\tau})^2 \sigma_{r,0}^2 + \gamma v^2(t/\tau) \right).
\end{equation}
In the case of an initially even community, $\sigma_{n,0} = 0$, and initial fitnesses drawn from the stationary distribution, $\sigma_{r,0} = \sigma_r$, we have 
\begin{equation}\label{eq:gtw}
    \gamma t_c w^2(t_c/\tau) = 2 \ln S,
\end{equation}
where
\begin{equation}
    w^2(s) := \frac{(1 - e^{-s})^2}{s} + v^2(s) = \frac{e^{-2s} - (1-2s)}{2s}.
\end{equation}
Using $w(s\gg 1) = 1$ and $w(s) = s + O(s^2)$,
\begin{equation}
    t_c = \begin{cases}
        \dfrac{2 \ln S}{\gamma}, & \tau \ll t_c, \\
        \dfrac{2 \ln S}{\gamma} \cdot \sqrt{ \frac{\gamma\tau}{2 \ln S}}, & \tau \gg t_c.\\ 
    \end{cases}
\end{equation}

Note, however, that the critical time should diverge in the large-community limit where the condensation phase transition becomes sharp; for finite communities the unevenness transition is therefore always observed as a smooth crossover of regimes. 

For a heuristic derivation of \eqref{eq:gtw}, following \cite{Derrida1981},  consider, at any given time, the variance in the normalization factor $Z$ in \eqref{eq:pi_sol} over many realizations of the fluctuating fitnesses: if it is large, that signals that the sum tends to be dominated by a few fluctuating terms. Denote $X_i(t) = \ln n_i(0) + R_i(t)$. The analysis is straightforward if the $X_i(t)$ are normal i.i.d. Then using $\avg{e^{X_i}} = \exp\{ \avg{X_i}+\Var{X}/2\}$,
\begin{equation}
    \avg{Z} = S e^{\avg{X} + \frac{1}{2}\Var{X}}, 
\end{equation}
\begin{equation}
    \avg{Z^2} = S e^{2\avg{X}}\left( e^{2\Var{X}} + (S-1)e^{\Var{X}}\right);
\end{equation}
hence, with $S-1 \approx S$,
\begin{equation}\label{eq:varZ}
    \frac{\Var Z}{\avg{Z}^2} = e^{\Var X - \ln S}.
\end{equation}
As $S$ becomes large, the $Z$-variance is non-negligible for times $t$ large enough that the above exponent is not very negative; $\Var{X(t)} \gtrsim \ln S$.

\section{Dimensional analysis}\label{sec:nondim}
Here we show which parameter combinations have qualitative effect on the model \eqref{eq:dotni} \& (\ref{eq:dotri}).

The special case \eqref{eq:neplicator} indicates that $1/\gamma$ (rather than $1/r^*$) is generally the natural timescale, and $K = r^*/\mu$ the natural scale of abundances. We therefore introduce the non-dimensional time $\tilde{t} := \gamma t$. We shift and rescale fitnesses as
$\tilde{r}_i(\tilde{t}) := (r_i(t) - r^*)/\gamma$, and define $\tilde{W}(\tilde{t}) \sim \mathcal{N}(0, \tilde{t})$, to transform \eqref{eq:dotri} into
\begin{equation}\label{eq:rtilde}
    (\tau\gamma) \deriv{\tilde{t}}\tilde{r}_i = - \tilde{r}_i + \deriv{\tilde{t}}\tilde{W}.
\end{equation}
We derive evolutions for $\tilde{N}(\tilde{t}) := N(t)/K$ and $\tilde{p}_i(\tilde{t}) := p_i(t)$, similar to the derivation of \eqref{eq:Ndot} and \eqref{eq:peplicator} in \autoref{app:replicator}, obtaining
\begin{equation}\label{eq:Ntilde}
    \deriv{\tilde{t}} \tilde{N} = \frac{r^*}{\gamma} \tilde{N}\left[ 1 - \left(1 + \frac{K \varepsilon}{r^*}  \frac{1}{\tilde{S}_\text{eff}}\right) \tilde{N} + \tilde{\rho} \right] + \frac{S\lambda}{\gamma K},
\end{equation}
and
\begin{multline}
    \deriv{\tilde{t}} \tilde{p}_i = \tilde{p}_i[\tilde{r}_i - \tilde{\rho}] + \frac{K \varepsilon}{\gamma} \tilde{N} \tilde{p}_i\left(\frac{1}{\tilde{S}_\text{eff}} - \tilde{p}_i\right) \\ + \frac{S \lambda}{\gamma K} \cdot \tilde{N}\left( \frac{1}{S} - \tilde{p}_i \right),\label{eq:ptilde}
\end{multline}
where $\tilde{\rho} := \sum_i \tilde{r_i}\tilde{p}_i$. As is to be expected, the set of equations (\ref{eq:rtilde})--(\ref{eq:ptilde}) depend (beside $S$) on the full set of non-dimensionalized parameters
\begin{equation}
    \tilde{\tau} := \gamma\tau,\quad\tilde{r}^* := \frac{r^*}{\gamma},\quad \tilde{\varepsilon} := \frac{K \varepsilon}{\gamma},\quad \tilde{\lambda}_\text{tot} := \frac{S\lambda}{\gamma K}.
\end{equation}
However, \eqref{eq:ptilde} only has a \textit{direct} dependence on $\tilde{\varepsilon}$ and $\tilde{\lambda}_\text{tot}$. So, if $\tilde{N}$ is relatively constant and $\tau$ not much larger than $1/\gamma$ we can expect the stationary distribution to depend mainly on $\tilde{\varepsilon}$ and $\tilde{\lambda}_{\text{tot}}$. To highlight their importance and interpretation we denote them instead by $\Sigma$ and $B$, \eqref{eq:BS}

As a complementary understanding of the reduced parameter space, consider rescaling time by $r^*$ (instead of $\gamma$) and define 
\begin{equation}\label{eq:altnondim}
    \hat{\gamma} := \frac{\gamma}{r^*},\quad \hat{\varepsilon} = \frac{K \varepsilon}{r^*},\quad \hat{\lambda}_\text{tot} = \frac{S\lambda}{Kr^*}.
\end{equation}
Then $\Sigma = \hat{\varepsilon}/\hat{\gamma}$ and $B=\hat{\lambda}_{\text{tot}}/\hat{\gamma}$. The plane $(B,\Sigma)$ can then be mapped into the simplex defined by the relative proportions of the compound parameters \eqref{eq:altnondim}:
\begin{subequations}
\begin{align}
    f_{\gamma} &:= \frac{\hat{\gamma}}{\hat{\gamma} + \hat{\varepsilon} +\hat{\lambda}_\text{tot} } = \frac{1}{1+\Sigma+B},\\
     f_{\epsilon} &:= \frac{\hat{\varepsilon} }{\hat{\gamma} + \hat{\varepsilon} +\hat{\lambda}_\text{tot} } = \frac{\Sigma}{1+\Sigma+B}, \\
      f_{\lambda} &:= \frac{\hat{\lambda}_\text{tot}}{\hat{\gamma} + \hat{\varepsilon} +\hat{\lambda}_\text{tot}}  = \frac{B}{1+\Sigma+B},
\end{align} 
\end{subequations}
and, conversely, $\Sigma = f_\varepsilon/f_\gamma$, $B = f_{\lambda}/f_\gamma$. That is, what matters for the shape of static abundance patterns (e.g.\ SAD) is essentially the relative strength of exclusion rate, excess self-limitation, and immigration.

\section{Stationary abundance distribution of the focal-species model}\label{sec:dist}

Matching the statistics of $r_\text{eff}(t)$ to $r_i(t) - \mu N(t)$ (for some arbitrary $i$), we set its mean to
\begin{equation}\label{eq:reff}
    r^*_\text{eff} = r^* - \mu \overline{N},
\end{equation}
with the over bar denoting an average over long times; the variance (neglecting the small covariance of focal species fitness and total abundance) to
\begin{equation}
    \sigma_{r_\text{eff}}^2 \approx \sigma_r^2 + \mu^2 \text{Var}[N];
\end{equation}
and $\tau_\text{eff} \approx \tau$, because fluctuations are chiefly driven by $r_i(t)$. Taking the noise statistics as given, the focal-species model \eqref{eq:focal} is a one-dimensional SDE whose stationary distribution can be obtained by standard techniques. When the noise is coloured, an approximate solution (see SI of \cite{Mallmin2024a} for derivation) is given by 
\begin{subequations}\label{eq:P(n)}
\begin{equation}
    P(n) \approx \frac{1}{\mathcal{N}}\left( \frac{1}{\tau_\text{eff}} + \varepsilon n + \frac{\lambda}{n} \right) n^{- \nu} \exp\left[ - q_+(\varepsilon n) - q_-(\lambda/n)  \right]
\end{equation}\label{eq:P(n)_a}
\begin{equation}
    \nu = 1 - \frac{2 r_\text{eff}^*}{\gamma_\text{eff}},\quad q_\pm(x) = \frac{\tau_{\text{eff}}}{\gamma_{\text{eff}}} (x + \tau_\text{eff}^{-1} \pm r^*_\text{eff})^2.
\end{equation}
\end{subequations}
In the fast-environment limit ($\tau \to 0$ at fixed $\gamma$), we obtain \eqref{eq:P(n)-lim} (which is the exact solution) by keeping from the large parenthesis in \eqref{eq:P(n)_a} only the constant diverging term, and expanding $q_\pm(x)$ to first finite order; apparent divergences of constants must cancel in the new normalization.

The normalization factor corresponding to \eqref{eq:P(n)-lim} is that of the generalized inverse Gaussian \cite{Jorgensen1982},
\begin{equation}
    \mathcal{N}_\text{fast-env} =  2 (ab)^{-\frac{1-\nu}{2}} \mathsf{K}_{1-\nu}\left(2\sqrt{b/a}\right),
\end{equation} 
where $\mathsf{K}_{p}(z)$ is the $p$-th order modified Bessel function of the second kind. The normalization factor $\mathcal{N}$ for \eqref{eq:P(n)_a} must be evaluated by numerical integration, and requires special care. Write \eqref{eq:P(n)_a} as 
\begin{equation}
    P(n) = e^{\mathcal{H}(n) - \ln \mathcal{N}},
\end{equation}
and evaluate $\ln \mathcal{N}$ according to 
\begin{equation}\label{eq:lnN}
    \ln \mathcal{N} = \mathcal{L}^* + \ln \int_{-\infty}^{\infty} \diff y\,e^{\mathcal{L}(y) - \mathcal{L}^*}
\end{equation}
where $\mathcal{L}(y) = \mathcal{H}(e^y) + y$ and $\mathcal{L}^* = \max_y \mathcal{L}(y)$. The integral appearing in \eqref{eq:lnN} can be evaluated with standard numerical integration.

In principle, the values of the effective parameters $r_\text{eff}^*$, $\gamma_\text{eff}$, $\tau_\text{eff}$ could be obtained through solving self-consistency relations, rather than being extracted from  simulation, at least for such values of the original parameters that fluctuations in $N(t)$ are small. 
To illustrate, consider the special case of no immigration and the fast-environment limit, and suppose $N(t)\approx \overline{N}$. The focal species then undergoes stochastic logistic growth and $P(n)$ is the PDF of the Gamma distribution with shape parameter $2r^*_\text{eff} / \gamma$ and scale parameter $\gamma/2\varepsilon$. On the one hand, since species fluctuate effectively independently, we posit 
\begin{equation}
     \overline{N} = S \int_0^\infty n P(n) \diff n,
\end{equation}
which evaluates to $S r^*_\text{eff}/{\varepsilon}$; on the other hand, we have the definition of $r^*_\text{eff}$ via $\overline{N}$, \eqref{eq:reff}. Combining these relations we find 
\begin{equation}
    r^*_\text{eff} = r^* \left( \frac{1}{1+ \frac{\mu S}{\varepsilon}} \right).
\end{equation}
The solution is valid (the Gamma distribution remains within its range of normalizability) as long as  $\varepsilon$ is positive. The solution is lost when $\epsilon \to 0$, as we indeed expect from our analysis of \eqref{eq:neplicator}.

\section{Coexistence fixed point for fixed fitness values}\label{app:fp}
We are solving for the fixed points of
\begin{equation}\label{eq:dotni-coex}
    \dot{n}_i(t) = n_i(t)[r_i - \mu N(t) - \varepsilon n_i(t)] = 0,\quad \varepsilon > 0,
\end{equation}
where $r_i$ are some fixed numbers. Let $s_i = 1$ if species $i$ survives in the fixed point and $0$ if it is extinct. Then its equilibrium abundance is
\begin{equation}\label{eq:nistar}
    \hat{n}_i = s_i \frac{r_i - \mu \hat{N}}{\varepsilon}, 
\end{equation}
where $\hat{N} = \sum_{j=1}^S \hat{n}_j$. Summing \eqref{eq:nistar} and solving for $\hat{N}$ yields 
\begin{equation}
    \hat{N} = \frac{\hat{r}}{\mu + \varepsilon/\hat{S}},
\end{equation}
where $\hat{S} = \sum_{j=1}^S s_j$ and $\hat{r} = \sum_{j=1}^S s_j r_j / \hat{S}$. Thus
\begin{equation}
    \hat{n}_i = s_i \frac{1}{\varepsilon}\left( r_i - \frac{\hat{r}}{1 + \frac{\varepsilon}{\mu \hat{S}}}\right).
\end{equation}
Feasibility requires that $r_i > \hat{r}/(1 + \varepsilon/\mu \hat{S})$ if $s_i =1$; uninvadability requires that $s_i = 1$ if this rate inequality holds. Stability of fixed points follows from the negative definiteness of the interaction matrix $[-\mu - \delta_{ij}\varepsilon]$; symmetric interactions preclude chaos or limit cycles \cite{Hofbauer2002}. Suppose $r_1 > r_2 > r_3 \ldots$. One concludes that there is a unique uninvadable fixed point consisting of the species with the largest $r_i$, up to the largest index $i$ for which $r_i > \sum_{j=1}^i r_j/(1 + \varepsilon/\mu i)$.

If the $r_i$ were drawn from $\mathcal{N}(r^*,\sigma_r)$, then the smallest among $S$ would be $r_S = r - h(S)\sigma_r$, where $h(S)$ is a random variable whose distribution for large $S$ is known from extreme value theory \cite{Vivo2015}. For $S=100,1000,10000$, $\avg{h(S)} \approx 2.5, 3.2, 3.9$, i.e.\ relatively insensitive to $S$. Thus, essentially all species coexist when 
\begin{equation}
    \frac{\varepsilon}{\mu} > S\left(\frac{1}{1 - h(S) \sigma_r / \hat{r}} -1\right) \approx 3 S \frac{\sigma_r}{r^*},
\end{equation}
if $\sigma_r/r^*$ is small.

\printbibliography

\end{document}


\maketitle

\FloatBarrier

\renewcommand*\listtablename{Supplementary Tables}
\listoftables 

\renewcommand*\listfigurename{Supplementary Figures}
\listoffigures

\vfill

\begin{table}[h!]

    \centering

    \begin{tabularx}{0.8\linewidth}{R{10mm}X}
        \hline
        \multicolumn{2}{c}{\textbf{Dynamical variables}}\\ 
        \hline
        $n_i(t)$ & abundance of species $i$ ($n_{i,\alpha}(t)$ for patch $\alpha$)\\
        $r_i(t)$ & intrinsic growth rate or environmental fitness---\textit{fitness}, for short---of species $i$ \\      
        \multicolumn{2}{p{\dimexpr0.8\linewidth-2\tabcolsep\relax}}{\it Derived quantities}\\
        $N(t)$ & $:= \sum_i n_i(t)$; total abundance \\
        $p_i(t)$ & $:= n_i(t)/N(t)$; relative abundance \\
        $\rho(t)$ & $:= \sum_i r_i(t)p_i(t)$; community-averaged fitness \\ 
        $S_\text{eff}(t)$ & $:= [\sum_i p_i^2(t)]^{-1}$; effective species richness, i.a.\ Simpson's (reciprocal) diversity index \\  
    \end{tabularx}

    \begin{tabularx}{0.8\linewidth}{R{10mm}XX}    
        \hline
        \multicolumn{3}{c}{\textbf{Base parameters}}\\ 
        \hline
        \multicolumn{3}{p{\dimexpr0.8\linewidth-2\tabcolsep\relax}}{\it The fundamental parameters that fully determine the model (note that \red{$\mu$} and \red{$\sigma_r^2$} are redundant): }\\
        $S$ & number of species & ---range 100--1000 \\
        \red{$\mu$} & heterospecific interaction rate &  ---implied by $r^*, K$; always 1 \\
        $\varepsilon$ & excess self-regulation & ---range 0.01--100 or zero \\
        $r^*$ & fitness mean value & ---fixed to 1 by non-dimensionalization \\
        $K$ & $:=r^*/\mu$; carrying capacity  & ---fixed to 1 by non-dimensionalization \\
        $\lambda$ & immigration rate & ---range $10^{-10}$--$10^{-4}$ or zero \\
        $\tau$ & autocorrelation time; \hfill\phantom{.} $\text{Corr}[r_i(t),r_j(t')] =  \delta_{ij} e^{-|t-t'|/\tau}$  &  ---default value 10; range $0.01$--$100$\\
        \red{$\sigma_r$} & std of fitness fluctuations & ---implied by $\tau,\gamma$; default value 0.05\\
        $\gamma$ & $:= 2\sigma_r^2 \tau$; rate of stochastic exclusion;  env.\ noise amplitude squared & ---default value 0.05; range $10^{-4}$--100\\
        $n_\text{ext}$ & extinction cutoff & ---range $10^{-12}$--$10^{-3}$ or absent (zero)\\ 
        \multicolumn{3}{p{\dimexpr0.8\linewidth-2\tabcolsep\relax}}{\it Instead of $\lambda$ in the multi-patch model:}\\
        $M$ & number of patches & ---range 1--80\\
        $d_{\beta \alpha}$ & rate of immigration from patch $\alpha$ to $\beta$& ---0.001 or $0.01/M$ \\
        \multicolumn{3}{p{\dimexpr0.8\linewidth-2\tabcolsep\relax}}{\it In the scenario without time-average neutrality:}\\
        $r_i^*$ & mean fitness of species $i$; drawn uniformly from $r^*\pm \delta r^*$ & ---maximal range 0.5--1.5
    \end{tabularx}
    
    \begin{tabularx}{0.8\linewidth}{R{10mm}X}
        \hline        
        \multicolumn{2}{c}{\textbf{Derived parameters}}\\ 
        \hline
        \multicolumn{2}{p{\dimexpr0.8\linewidth-2\tabcolsep\relax}}{\it Compound parameters defining the Buffering--Stabilization plane}\\
        $B$ & $:= S\lambda/K\gamma$; Buffering \\
        $\Sigma$ & $:= K\varepsilon/\gamma$; Stabilization\\
        \multicolumn{2}{p{\dimexpr0.8\linewidth-2\tabcolsep\relax}}{\it `Effective parameters' determined implicitly by the community dynamics as the parameters of the OUP approximation of $r_i(t) - \mu N(t)$}\\
        $r_\text{eff}^*$ & effective mean fitness \\
        $\sigma_{r_\text{eff}}$ & effective fitness std\\
        $\tau_{\text{eff}}$ & effective fitness autocorrelation time \\
        \multicolumn{2}{p{\dimexpr0.8\linewidth-2\tabcolsep\relax}}{\it Parameters of the GIG distribution describing the SAD and FAD under TAN}\\
        $a$ & $:=  \gamma_\text{eff}/2 \varepsilon$; characteristic abundance of right bend \\
        $b$ & $:= 2 \lambda/ \gamma_\text{eff}$; characteristic abundance of left bend \\
        $\nu$ & $:=  1 - 2 r^*_\text{eff}/\gamma_\text{eff}$; exponent of the (inverse) power law section
    \end{tabularx}
    \begin{tabularx}{0.8\linewidth}{R{10mm}X}
        \hline
        \multicolumn{2}{c}{\textbf{Abbreviations}}\\ 
        \hline
        TAN & Time-average neutrality; species have identical expected fitness \\
        SAD & Species--abundance distribution; fraction of species vs abundance class \\
        FAD & Frequency--abundance distribution; fraction of time spent in abundance class by one particular species \\
        GIG & generalized inverse Gaussian (distribution); a three-parameter `bent' power law
    \end{tabularx}
    \mycaption{Table of variables, parameters, and abbreviations}{The parameter values used in simulation are stated in the corresponding figure captions. Here, we give an indication of the default values and/or ranges considered.}\label{tab:symb}
\end{table}

\begin{figure}[h!]
    \centering
    \includegraphics[]{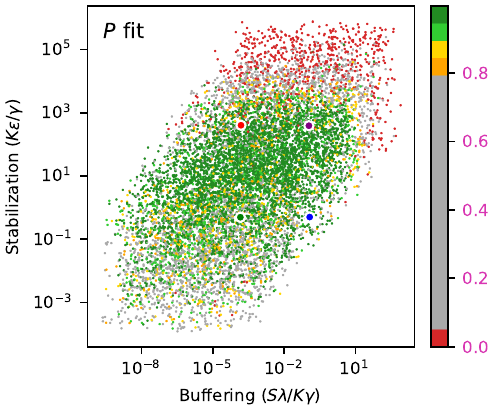}
	\mycaption{Goodness of fit for the focal species model SAD prediction given simulated noise statistics}{For each simulation we have produced $P(n)$ according to \maineqref{eq:P(n)-lim}{17}, with the parameters $\nu, a, b$ obtained from \maineqref{eq:nu}{18} and (\mref{eq:ab}{19}) using the values of $\overline{N}$ and $\text{Var}[N]$ from the simulations. The goodness of fit of $P(n)$ with the time-averaged FAD (from 0 (no match) to 1 (perfect match)) is measured by (one minus) the Kolmogorov-Smirnov distance of the distributions:  $1 - \sup_n |F_P(n) - F_\text{FAD}(n)|$. }\label{fig:Pfit}
\end{figure}

\begin{figure}[h!]
    \centering
    \includegraphics[]{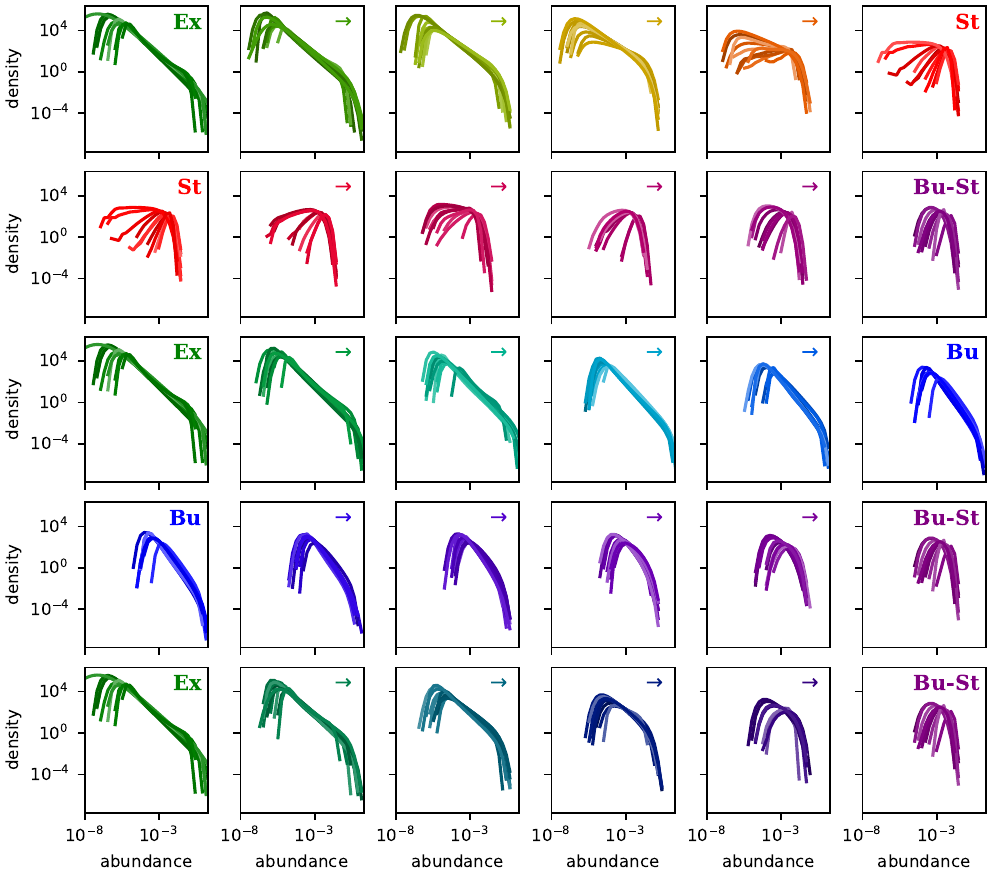}
	\mycaption{Systematic change in SAD shape by traversing the Buffering-Stabilization parameter plane}{ (With reference to \mainfigref{fig:distributions}{5}.) Here we show the distribution sets that lie on a straight line between the reference cases, e.g.\ between Ex and St, in the upper row; Ex and Bu-St in the second row; and so on. The variation in line colors within a panel are just a guide for the eye.}\label{fig:fad_grid}
\end{figure}

\begin{figure}[h!]
    \centering
    \includegraphics[]{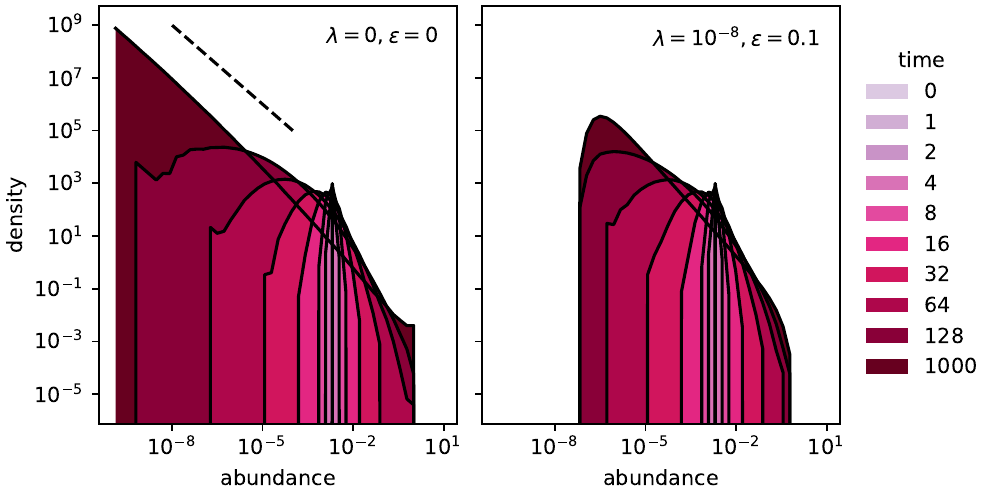}
	\caption[Time-evolution of the SAD with or without coexistence mechanisms]{\textbf{Time-evolution of the SAD with or without coexistence mechanisms.} Starting from an even initial condition, we track the evolution of the SAD averaged over 1000 realizations. In the left panel, there is no immigration and no additional self-suppression, in contrast to the right panel. Early on, both scenarios give similar distributions, until the bounds in the latter scenario restrict the expansion of the distribution; for the former scenario, the power-law section extends ever into lower abundance decades with time, seemingly approaching an exponent of 1 (dashed line). Here, $S=500, r^*=1, \sigma_r= 0.05, \tau=10$.}\label{fig:sad_spread}
\end{figure}

\begin{figure}[h!]
    \centering
    \includegraphics[]{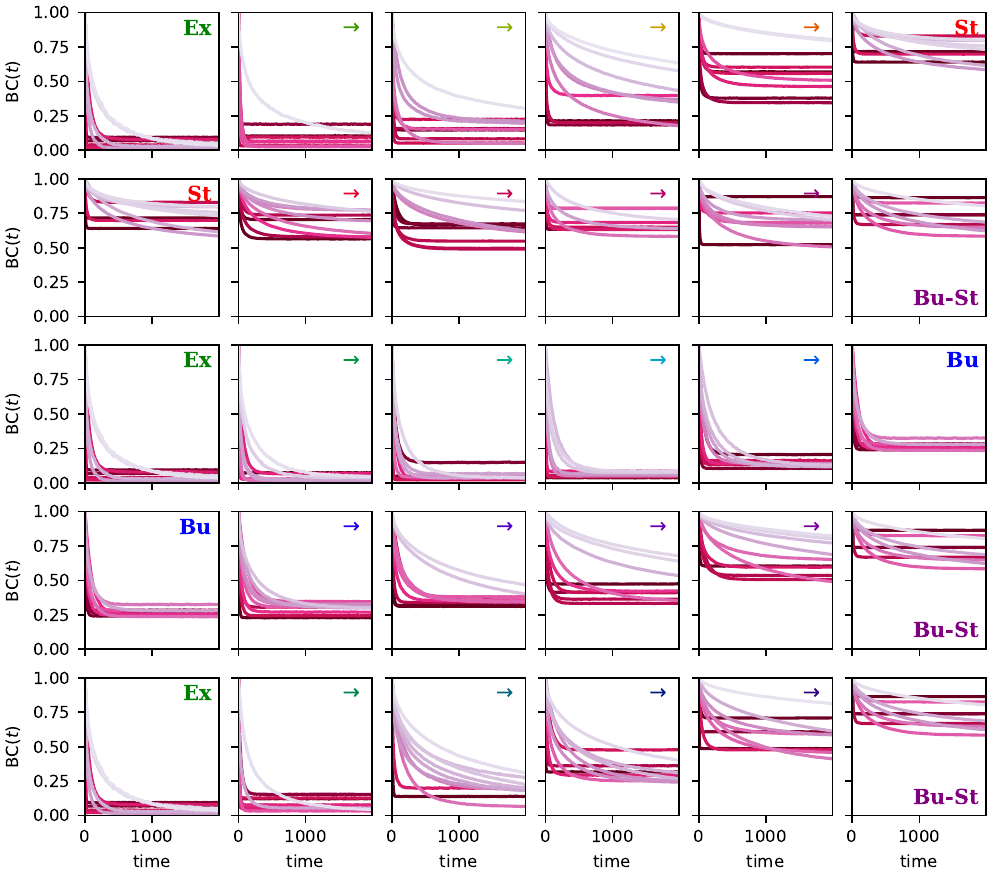}
	\mycaption{Turnover measured by Bray-Curtis decay corresponding to the panels in \autoref{fig:fad_grid}}{The color of the line reflects the value of $\log_{10}\gamma$, normalized separately for each panel---light colors for small $\gamma$, dark color for large. Note that small $\gamma$ gives slower decay, and that narrow SADs are associated with high limit of the BC.}\label{fig:bc_grid}
\end{figure}

\begin{figure}[h!]
    \centering
    \includegraphics[]{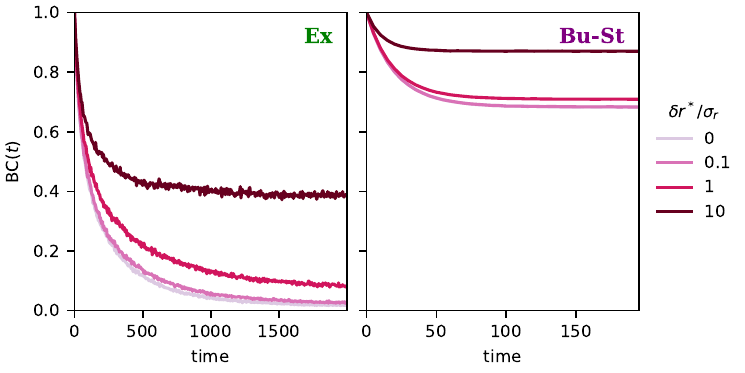}
	\mycaption{Turnover measured by Bray-Curtis decay corresponding to \mainfigref{fig:nontan}{6}}{For each scenario (row) of \mainfigref{fig:nontan}{6}, each of the four panels with different $\delta r^*/\sigma_r$ corresponds to one line. Note the 10x difference in timescale between the two scenarios.} \label{fig:bc6}
\end{figure}

\begin{figure}[h!]
    \centering
    \includegraphics[]{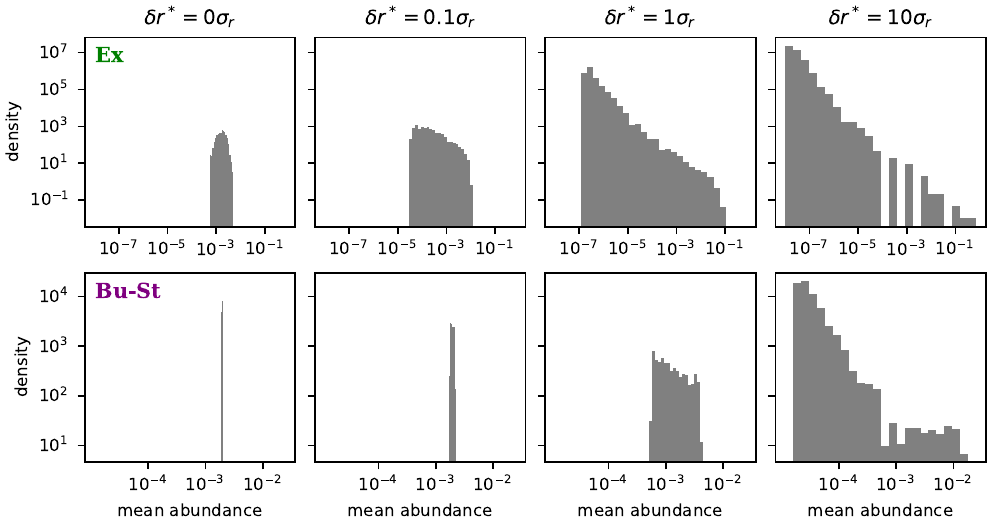}
	\mycaption{Distribution of mean abundances for the panels in \mainfigref{fig:nontan}{6}}{ Breaking time-averaged neutrality produces a distribution of species means. Note that the panels of the first column are TAN, so theoretically all species would have the same mean if the sampling time window was infinite. Note also the difference in scale of the horizontal axis of the two rows.}\label{fig:mean_dist}
\end{figure}

\FloatBarrier